\documentclass[prd,aps,nofootinbib,notitlepage,onecolumn,superscriptaddress]{revtex4-1}

\usepackage{comment,verbatim}
\usepackage[sort&compress]{natbib}

\usepackage[a4paper,top=2cm,bottom=2cm,left=3cm,right=3cm,marginparwidth=1.75cm]{geometry}

\usepackage{xcolor}
\usepackage{amsmath}
\usepackage{graphicx}

\usepackage{pifont}
\newcommand{\cmark}{\ding{51}}%
\newcommand{\xmark}{\ding{55}}%

\usepackage[colorlinks=true, allcolors=blue]{hyperref}

\newcommand{\re}{\text{Re}\,}
\newcommand{\im}{\text{Im}\,}

\newcommand{\anna}[1]{\textcolor{blue}{[\textbf{Anna:} #1]}}
\newcommand{\merce}[1]{\textcolor{violet}{[\textbf{Mercè:} #1]}}

\usepackage{booktabs}

\date{}
\begin{abstract}
Observations of energy-dependent photon time delays from distant flaring sources provide significant constraints on Lorentz Invariance Violation (LIV). Such effects originate from modified vacuum dispersion relations, causing differences in propagation times
for photons emitted simultaneously from gamma-ray bursts, active galactic nuclei, or pulsars. These modifications are often parametrized within a general framework by an effective quantum gravity energy scale $E_{QG,n}$. While such general constraints are well established in the LIV literature, their translation into specific coefficients of alternative theoretical frameworks, such as the Standard-Model Extension (SME), is rarely carried out. In particular, existing bounds on the quadratic case ($n=2$) of $E_{QG,n}$ can be systematically converted into constraints on the non-birefringent, CPT-conserving SME coefficients $c^{(6)}_{(I)jm}$. This work provides a concise overview of the relevant SME formalism and introduces a transparent conversion method from $E_{QG,2}$ to SME parameters. We review the most stringent time-of-flight-based bounds on $E_{QG,n}$ and standardize them by accounting for systematics, applying missing prefactors, and transforming results into two-sided Gaussian uncertainties where needed. We then use these standardized constraints, along with additional bounds from the literature, to improve bounds on the individual SME coefficients of the photon sector by about an order of magnitude. A consistent methodology is developed to perform this conversion from the general LIV framework to the SME formalism.

\end{abstract}

\begin{document}

\title{Bounding anisotropic Lorentz Invariance Violation from measurements of the effective energy scale of quantum gravity}

\author{Merce Guerrero}
\email{merce.guerrero@ua.pt}
\affiliation{Departamento de Matematica da Universidade de Aveiro, Campus de Santiago, 3810-193 Aveiro, Portugal.}
\affiliation{Center for Research and Development in Mathematics and Applications (CIDMA), Campus de Santiago, 3810-193 Aveiro, Portugal.}

\author{Anna Campoy-Ordaz}
\email{campoy@ieec.cat}
\affiliation{{Departament de F\'{i}sica, Universitat Aut\`{o}noma de Barcelona and CERES-IEEC, 08193 Bellaterra, Spain}}

\author{Robertus Potting}
\email{rpotting@ualg.pt}
\affiliation{Center for Research and Development in Mathematics and Applications (CIDMA), Campus de Santiago, 3810-193 Aveiro, Portugal.}
\affiliation{Universidade do Algarve, Faculdade de Ci\^{e}ncias e Tecnologia, 8005-139 Faro, Portugal}

\author{Markus Gaug}
\email{markus.gaug@uab.cat}
\affiliation{{Departament de F\'{i}sica, Universitat Aut\`{o}noma de Barcelona and CERES-IEEC, 08193 Bellaterra, Spain}}

\maketitle

\setlength{\heavyrulewidth}{0.08em}   
\setlength{\lightrulewidth}{0.05em}   
\setlength{\cmidrulewidth}{0.03em}    

\section{Introduction}

The most stringent bounds on a modified dispersion relation of the photon in vacuum have been derived from astrophysical tests of energy-dependent time delays of photons from very distant but fast-varying (flaring) sources. In this scenario, simultaneously emitted photons of different energies would be detected on Earth at different times, due to possible Lorentz-Invariance-Violating (LIV) properties of the vacuum~\citep{Amelino_Camelia:1998,Mattingly_2005,AmelinoCamelia:2013,Addazi:2022,Bolmont:2022}.  
In the most generic approach, 
the modified vacuum dispersion relation for the photon is expanded into
a power-law series  
of the following
form~\citep{Amelino_Camelia:1998}:

\begin{equation}\label{Eq: Energy DR}
    E^2 \simeq p^2c^2 \times \left[1-\sum_{n=1}^{\infty} s_\pm \left( \dfrac{E}{E_{QG,n}}\right)^n\right].
\end{equation}
Here, $c$ is the speed of light, $s_\pm$ defines the sign of the (LIV) effect ($s_\pm=-1$ for superluminal and $s_\pm=+1$ for subluminal behaviour) and $E_{QG,n}$ refers to an effective energy scale of quantum gravity~\citep{Amelino_Camelia:1998}. The parameter $n$ denotes the expansion order, i.e., $n=1$ is used for a deviation from the trivial case that scales linearly with energy, while $n=2$ is for the quadratic case. From Eq.~\eqref{Eq: Energy DR},  the photon's group velocity can be obtained assuming the dominant lowest-order term is $n$:
\begin{equation}
    u_\textit{ph}(E)=\dfrac{\partial E}{\partial p} \simeq c \times \left[1- s_\pm \dfrac{n+1}{2}\left( \dfrac{E}{E_{QG,n}}\right)^n\right].
    \label{eq:group_velocity_liv}
\end{equation}

Hence, two photons with different observed energies $E_h$ and $E_l$ ($E_h>E_l$), emitted simultaneously from the same source, will reach Earth with a time delay, which for large distances depends on the cosmological model. In its first version presented~\citep{JacobPiran:2008}, the time delay is expressed as: 

\begin{equation}
\label{eq:time_delay_LIV}
\Delta t \simeq (E_{h}^n - E_{l}^n) \mathit{s_\pm} \frac{(n+1)}{2H_0}\frac{1}{E_{QG,n}^n}\times \int _0^z\frac{(1+z')^n}{\sqrt{\Omega_{\rm\Lambda} + \Omega_{\rm M} (1 + z')^3}}dz' \ ,
\end{equation}
\noindent
but other, more complicated formulations of the redshift dependence have been proposed~\citep{Pfeifer:2018,caroff2024discriminatingdifferentmodifieddispersion}. 

bounds on $E_\textit{QG,1}$ up to and well beyond the Planck energy have been obtained with this method from Gamma-Ray-Bursts (GRBs)~\citep{Boggs:2004,Abdo:2009,Vasileiou:2013,Ellis:2019,GRB1901114LIV:2020,lhaaso_2024} and Active Galactic Nuclei (AGNs)~\citep{Mkn501LIV:2009,PKS2155LIV:2011,Kislat:2015,Mrk421LIV:2024}. 
Furthermore, the same sources and pulsars~\citep{PulsarLIV:2017} have contributed to strong bounds on $E_\textit{QG,2}$, although those are still orders of magnitude below the Planck energy scale.

Moving to a more specific case, various scenarios have been formulated that may result in the generation of a LIV-induced vacuum dispersion relation for photons. It is important to note that all these scenarios assume quantum properties of gravity~\citep{Kempf:1995,Gambini:1999,Alfaro:2002,Ellis:2008,Das:2010,Amelino-Camelia:2011}. 

In a complementary approach, proposed by Kosteleck{\'y} and collaborators, violations of Lorentz symmetry at attainable energies are described by the Standard-Model Extension (SME)~\citep{Colladay:1996iz, Colladay:1998fq, Kostelecky:2003fs}. 
This is an effective field theory allowing for the possibility of Lorentz and CPT violation in the Standard Model of particles coupled to General Relativity. 
The SME is generally chosen to preserve energy-momentum conservation, observer Lorentz invariance, hermiticity, micro-causality, gauge invariance and reparametrization invariance.
Each Lorentz-violating term in the Lagrangian density of the SME is an observer scalar density formed by contracting a Lorentz-violating operator of a certain mass dimension with a coefficient governing the size of the effect.
These can, in principle, be measured by appropriate experiments. 
All possible SME terms affecting photon propagation have been constructed explicitly~\citep{Kostelecky:2009zp}.

A large, ever-expanding set of experimental bounds on the SME coefficients has been published.
An overview can be found in a set of data tables elaborated by \citet{Kostelecky:2011}\footnote{They are updated on a yearly basis on a pre-print server (see \url{https://arxiv.org/abs/0801.0287}).}.
However, it is useful to note that publications leading to entries in these tables are not independently checked for quality of the methodology and analysis, as is clearly stated in the introduction
of the tables. At the same time, the astrophysical LIV community rarely converts their bounds on $E_{QG,n}$ to those that match the SME framework (with the commendable exceptions of~\cite{Vasileiou:2013,Kislat:2015,Wei:2022}).
In particular, the bounds on $E_{QG,2}$ could and should also be converted to the non-birefringent and CPT-conserving coupling SME coefficients. 

The purpose of this paper is to provide an overview of the SME formalism leading to an easy and straightforward conversion recipe from $E_{QG,2}$ to SME coefficients. In this process, we critically review publications in which such a conversion has been made, including the data tables~\citep{Kostelecky:2011}. We will correct inconsistencies introduced during early papers on $E_{QG,n}$ and present a consistent collection of bounds on the SME coefficients $c^{(6)}_{(I)jm}$, which differ from the current version of the data tables and extend those. Finally, we provide a simple new recipe to deconvolute a combination of bounds on $E_{QG,2}$ obtained from sources at different directions into a set of bounds on the individual SME coefficients.






The article is organised as follows: Section~\ref{sec:theory} provides a review of the SME formalism in the photon sector, while Section~\ref{Sec: Disp test} outlines the procedure to convert constraints on the quantum gravity scale $E_{QG,2}$ into bounds on SME coefficients. In Section~\ref{sec:results}, a compilation and critical comparison of the most stringent existing bounds, including necessary corrections, is presented; also, individual constraints on the SME coefficients $k_{(I)jm}^{(6)}$ are derived through a statistical combination from multiple astrophysical sources. Finally, Section~\ref{sec:conclusions} draws the conclusions. 

\section{The photon sector of the Standard-Model Extension}
\label{sec:theory}

Numerous studies have proposed that one of the physical effects of quantum gravity may be the violation of Lorentz invariance \citep{KosteleckySamuel_1989,KosteleckyPotting_1991,Kostelecky:1994rn,Mattingly_2005,AmelinoCamelia:2013,Mavromatos_2010,Ellis:2011ek,Tasson:2014dfa}.
One of the most useful frameworks for studying these effects is the Standard-Model Extension (SME)~\citep{Colladay:1998fq,Kostelecky:2003fs,Bluhm_2006}.
In this work, we are interested in the photon sector developed in detail in a series of works \cite{Colladay:1998fq, Kostelecky_2001, Kostelecky:2002hh, Kostelecky:2009zp}.
In this section, we will take the opportunity to present a short review of the most relevant aspects and formulae.
For a more complete treatment, we refer to the relevant literature,
in particular to~\citep{Kostelecky:2009zp}.

The pure photon sector of the SME is entirely described by the Lagrangian 
\begin{equation}
\label{Eq: Lagrangian-density}
\mathcal{L}= -\dfrac{1}{4} F_{\mu\nu}F^{\mu\nu}
             - \dfrac{1}{4} F_{\kappa\lambda}(\hat{k}_F)^{\kappa\lambda\mu\nu}F_{\mu\nu}
             + \dfrac{1}{2} \epsilon^{\kappa\lambda\mu\nu}A_\lambda(\hat{k}_{AF})_\kappa F_{\mu\nu} \ ,
\end{equation}
where $A_\mu$ is the usual magnetic vector potential while $F_{\mu\nu}$ is the field-strength tensor.
The operator-valued coefficients in the second and third terms on the right-hand side are defined by \citep{Kostelecky:2009zp}
\begin{align}
 \label{Eq:kF}
    (\hat{k}_F)^{\kappa\lambda\mu\nu} &= \sum_{d=\mbox{\footnotesize even}}(k_F^{(d)})^{\kappa\lambda\mu\nu\alpha_1...\alpha_{d-4}}\partial_{\alpha_1} ... \partial_{\alpha_{d-4}}~, \\
    (\hat{k}_{AF})_\kappa &= \sum_{d=\mbox{\footnotesize odd}}(k_{AF}^{(d)})_\kappa{}^{\alpha_1...\alpha_{d-3}}\partial_{\alpha_1} ... \partial_{\alpha_{d-3}}~,
  \label{Eq:kAF}
\end{align}
with the mass dimension $d\geq 3$, defined by partial spacetime derivatives contracted with the (constant) coefficients 
$(k^{(d)}_{F})^{\kappa\lambda\mu\nu\alpha_1...\alpha_{d-4}}$ and
$(k_{AF}^{(d)})_\kappa{}^{\alpha_1...\alpha_{d-3}}$.
The latter parametrize, respectively, the most general CPT-even and CPT-odd operators in the photon sector of the SME with mass dimension $d$.\footnote{Silently, it is assumed here that we are working in Minkowski space. In fact, we will be considering Friedmann-Lemaitre-Robertson-Walker (FLRW) type cosmological spacetimes. Strictly, in such a curved-space context, one should consider a more general expansion involving covariant derivatives and contractions, as well as the Riemann tensor. Such a more general framework for the SME has been developed recently~\citep{KosteleckyLi:2020}. However, the relevant curved-space corrections, which will involve powers of the Hubble constant, are extremely tiny compared to the photon momenta we will be considering in this work. Therefore, the Lagrangian density (Eq.~\eqref{Eq: Lagrangian-density}) is perfectly adequate for our purposes.}

Of prime interest is the photon dispersion relation, which follows from the equations of motion corresponding to the Lagrangian density Eq.~\eqref{Eq: Lagrangian-density}.
Adopting the ansatz
\begin{equation} \label{Eq: ansatz}
    A_\mu (x) = A_\mu (p) e^{-ix\cdot p}~,
\end{equation}
the equation of motion can be expressed as
\begin{equation}\label{Eq: field eq M}
     M^{\mu\nu} A_\nu=0 \ ,
\end{equation}
with
\begin{equation}
   M^{\mu\nu} = \left(\eta^{\mu\nu}\eta^{\alpha\beta}-\eta^{\mu\alpha}\eta^{\nu\beta} + 2(\hat k_F)^{\mu\alpha\nu\beta}\right) p_\alpha p_\beta
   - 2i\epsilon^{\mu\nu\alpha\beta}(\hat k_{AF})_\alpha p_\beta \ ,
   \label{eq:M-mu-nu}
\end{equation}
where it is understood that each occurrence of $\partial_\rho$ in the
operators $\hat k_{AF}$ and $\hat k_F$ is replaced by $-i p_\rho$.
The CPT-violating coefficients $(\hat k_{AF})_\alpha$ lead to photon birefringence  \citep{Kostelecky:2009zp}.
The bounds obtained from the absence of vacuum birefringence via analogous sources of X-ray and gamma-ray polarimetry~\citep{Kislat:2017,Wei:2022} are by several orders of magnitude more stringent than those from a modified dispersion relation. For this reason, the present study will focus exclusively on non-birefringent photon propagation, thereby excluding the other terms from the subsequent analysis.


The coefficient $(\hat k_F)^{\mu\alpha\nu\beta}$ has the symmetries of the Riemann tensor.
It can therefore be expanded in a Weyl decomposition as
\begin{equation}
(\hat k_F)^{\mu\alpha\nu\beta} = \tfrac{1}{2}\left(\eta^{\mu\nu}\hat{c}_F^{\alpha\beta} + 
\eta^{\alpha\beta}\hat{c}_F^{\mu\nu} - \eta^{\nu\alpha}\hat{c}_F^{\mu\beta}
 - \eta^{\mu\beta}\hat{c}_F^{\nu\alpha}\right) + \hat C^{\mu\alpha\nu\beta}
 \label{eq:Weyl-domposition}
\end{equation}
where the tensor $\hat C^{\mu\alpha\nu\beta}$ corresponds to the Weyl component satisfying
$\hat C^{\mu\alpha\nu\beta}\eta_{\alpha\beta} = 0$,
while the terms corresponding to the Ricci component involve the tensor
\begin{equation}
\hat{c}_F^{\alpha\beta} = (\hat{k}_F)^{\alpha\lambda\beta}{}_\lambda - \tfrac{1}{6}\eta^{\alpha\beta}(\hat{k}_F)^{\mu\nu}{}_{\mu\nu} .
\label{eq:cF_kF}
\end{equation}
The Weyl component leads to birefringence \cite{Kostelecky:2009zp},
while (at leading order) the Ricci component does not. Therefore, we will consider, in the following, only the case $\hat C^{\mu\alpha\nu\beta} = 0$.

To first order in the coefficients $\hat{c}_F^{\alpha\beta}$,
the equation of motion \eqref{Eq: field eq M}, can be written as
\begin{equation}
\left(\bigl(\hat{g}^{-1}\bigr)^{\mu\nu}\bigl(\hat{g}^{-1}\bigr)^{\alpha\beta} -
\bigl(\hat{g}^{-1}\bigr)^{\mu\alpha}\bigl(\hat{g}^{-1}\bigr)^{\nu\beta}\right)p_\alpha p_\beta A_\nu = 0
\label{eq:field-eq2}
\end{equation}
where we introduced the ``inverse effective metric''
\begin{equation}
\bigl(\hat{g}^{-1}\bigr)^{\mu\nu} = \eta^{\mu\nu} + \hat{c}_F^{\mu\nu}\>.
\label{eq:inverse-effective-metric}
\end{equation}
The equation of motion \eqref{eq:field-eq2} is invariant under gauge transformations
$A_\mu \to A_\mu + \epsilon p_\mu$,
allowing to impose the ``deformed Lorenz gauge'' 
$p_\mu\bigl(\hat{g}^{-1}\bigr)^{\mu\nu} A_\nu = 0$.
The equation of motion \eqref{eq:field-eq2} then reduces to
\begin{equation}
p_\alpha\bigl(\hat{g}^{-1}\bigr)^{\alpha\beta}p_\beta\,\bigl(\hat{g}^{-1}\bigr)^{\mu\nu}A_\nu = 0
\end{equation}
from which we immediately deduce the dispersion relation
\begin{equation}
p_\alpha\bigl(\hat{g}^{-1}\bigr)^{\alpha\beta}p_\beta = p^\mu p_\mu + \hat{c}_F^{\mu\nu}p_\mu p_\nu = 0
\label{eq:dispersion-relation}
\end{equation}

when working to leading order in the Lorentz-violating coefficients,
we are free to replace $p^0$ in $\hat{c}_F^{\mu\nu}p_\mu p_\nu$
by its vacuum plane-wave value $\omega = p \equiv |\vec{p}|$ \cite{Kostelecky:2002hh,Kostelecky:2009zp}
(note that the coefficients $\hat{c}_F^{\mu\nu}$ themselves are momentum dependent).
The dispersion relation \eqref{eq:dispersion-relation} can then be expressed as
\begin{equation}
p^0 = (1 - \varsigma^0)p
\label{eq:dispersion-relation2}
\end{equation}
with 
%
\begin{equation}
\varsigma^0 = \tfrac{1}{2}\bigl(\hat{c}_F\bigr)^{\mu\nu}p_\mu p_\nu/\omega^2.
\label{eq:varsigma0}
\end{equation}
Note that we adopted the metric signature $(+---)$. 
An analysis that also takes into account the Weyl part of the coefficients $\hat k_F$ and the coefficients $\hat{k}_{AF}$ yields a generalization of relations \eqref{eq:dispersion-relation} and \eqref{eq:dispersion-relation2}, see \cite{Kostelecky:2008be}

It is useful to note that the coefficients $\bigl(\hat{c}_F\bigr)^{\mu\nu}$
are momentum dependent and can be expanded in accordance with Eqs.\ \eqref{eq:cF_kF} and \eqref{Eq:kF}:
%
\begin{equation}
\hat{c}_F^{\mu\nu}p_\mu p_\nu = \sum_{d=\mbox{\footnotesize even},d\ge4}\bigl(\tilde c_F^{(d)}\bigr)^{\mu\nu\alpha_1\ldots\alpha_{d-4}}p_\mu p_\nu p_{\alpha_1}\ldots p_{\alpha_{d-4}}
\label{eq:c_F_expansion}
\end{equation}
The structure of Eq.~\eqref{eq:c_F_expansion} shows that the coefficients $\bigl(\tilde c_F^{(d)}\bigr)^{\mu\nu\alpha_1\ldots\alpha_{d-4}}$ can be taken
totally symmetric in all its indices.
This can be enforced by imposing the constraint
\begin{equation}
\partial_{\vphantom{F}p}^{[\mu}\hat{c}_F^{\rho]\nu} = 0\>,
\label{eq:antisymmetrization}
\end{equation}
 where $\partial_p^\mu \equiv \partial/\partial p_\mu$ and the square brackets denote anti-symmetrization. 
A natural way to satisfy condition~\eqref{eq:antisymmetrization} is by introducing a scalar potential $\hat\Phi_F$
and taking
\begin{equation}
\hat{c}_F^{\mu \nu} = \partial_p^{\mu} \partial_p^{\nu} \hat\Phi_F
\label{eq:c_F-Phi_F}
\end{equation}
It is then convenient to expand $\hat\Phi_F$ in spherical harmonics,
\begin{equation}
   \hat\Phi_F = \sum_{d,n,j,m} \omega^{d-2-n} p^n\,{}_{0}Y_{jm}(\hat{p})
   (c_F^{(d)})_{njm}^{(0E)}\ .
\label{eq:Phi_F-spherical}
\end{equation}
Here, the index ranges satisfy $d = 4,6,8,\ldots$; $0\le n\le d-2$;
$j=n, n-2, n-4, ...$ ($j\ge 0$); and $-j\le m\le j$.
Note that the expansion \eqref{eq:Phi_F-spherical} does not assume
(yet) the leading-order approximation $\omega = p$.
It then follows from Eqs.~\eqref{eq:varsigma0} and \eqref{eq:c_F-Phi_F}, after some algebra, that
\begin{equation}
\varsigma^0 = \sum_{d,j,m}\omega^{d-4}(-1)^j\,{}_{0}Y_{jm}(\hat{p})
c^{(d)}_{(I)jm}
\end{equation}
where
\begin{equation}
c_{(I)jm}^{(d)} = \frac{1}{2} (d-2)(d-3)(-1)^j \sum_n (c_F^{(d)})_{njm}^{(0E)}
\label{eq:c_I}
\end{equation}
are a set of $(d-1)^2$ coefficients parametrizing the non-birefringent part at dimension $d$ of the $\hat k_F^{\mu\alpha\nu\beta}$ coefficients.
Note that in Eq.\ \eqref{eq:c_I} we imposed the leading-order ``vacuum-model'' approximation $\omega=p$ for the terms on the right-hand side \citep{Kostelecky:2009zp}.

Sometimes, the vacuum model is restricted to the isotropic subset of LIV operators that are invariant under spatial rotations. Note that the notion of spatial rotations (as opposed to boosts) is frame-dependent; in other words, it assumes the presence of a preferred observer Lorentz frame.
This model is obtained by setting all coefficients to 0, except for those with $j=0$, which means that the total angular momentum is 0. Therefore, the only non-zero coefficient that we are interested in for our case is
\begin{equation}
    (\mathring{c}_F^{(d)})_n =(c_F^{(d)})^{(0E)}_{n00}\ ,
\end{equation}
since it controls the non-birefringent operators in the preferred frame. The index $n$ determines the wavelength or frequency dependence, while the other two indices correspond to $j=0$ and $m=0$, the total angular momentum and its $z$-component, respectively.  The superscript $0$ refers to the spin weight of the operator, and with $E$ parity, that is, it transforms as $(-1)^{j+1}$ under parity.

\section{Dispersion tests}
\label{Sec: Disp test}

In this section, we explore how time delays between high-energy photons from distant astrophysical sources can be used to test for possible violations of Lorentz invariance. If photons of different energies travel at slightly different speeds, even small differences can lead to observable delays over cosmological distances. These delays provide a way to probe modified dispersion relations and place constraints on SME coefficients. We show how previous bounds on $E_{QG,n}$ can be translated into bounds on SME coefficients.

Assuming that the temporal emission of the source is well understood, we can use the time delays to constrain the velocities depending on the photon energies. Setting to zero all the birefringent coefficients, the group velocity of the photon becomes
\begin{equation}
    v_{gr} \simeq  1- (d-3)\varsigma^0 = 1- (d-3)
    \sum_{djm}
E^{d-4} \, _{0}Y_{jm}({\hat n})\, c^{(d)}_{(I)jm} .
\end{equation}
As demonstrated by the above equation, the minimum SME, which only includes $d=4$, is not enough to test a modified dispersion relation, since there is no difference in velocity for different energies. Consequently, considering higher-dimensional operators becomes imperative.

Assuming that the photons are emitted simultaneously by the same source, the co-moving distance between the source and the Earth is the same for both photons. However, due to the difference in group velocity, two photons observed with energies $E_l$ and $E_h$ will have a time delay given by~\citep{Kostelecky:2008be} 
\begin{equation}
    \Delta t \approx (d-3)(E^{d-4}_h -E^{d-4}_l)\int_0^z \dfrac{(1+z)^{d-4}}{H_z}dz \sum_{jm} {}_{0}Y_{jm} c^{(d)}_{(I)jm}
    \label{eq:time_delay_SME}
\end{equation}
where $H_z$ is the Hubble expansion rate at redshift $z$ defined as\footnote{Note that $\Omega_r$ and $\Omega_k$ have been frequently neglected~\citep{JacobPiran:2008,Pfeifer:2018,caroff2024discriminatingdifferentmodifieddispersion} due to their relative smallness~\citep{Navas:2025}.}
\begin{equation}
    H_z = H_0[\Omega_r(1+z)^4+\Omega_M(1+z)^3+\Omega_k (1+z)^2 +\Omega_\Lambda]^{1/2} \ .
\end{equation}

Hence, by comparing Eq.~\ref{eq:time_delay_LIV} with Eq.~\ref{eq:time_delay_SME}, one can achieve the objective of this work: to establish a straightforward conversion between the general LIV parameters and the SME coefficients, leading to the following formal substitution:
\begin{equation}
    \frac{s_\pm}{2 E^{d-4}_{QG,d-4}} \rightarrow \sum_{jm} Y_{jm} c^{(d)}_{(I)jm}
    \label{eq:relation_SME_LIV}
\end{equation}

whereby convention $s_\pm=+1$ would apply for a positive sum on the right side and $s_\pm=-1$ for a negative sum. Recalling that the corresponding SME coefficient of $E_{QG,1}$ is birefringent, we will only focus on the cases of $(d-4)$ even and greater than or equal to 2. The bounds treated in this article apply to the particular case of $E_{QG,2}$.

\section{Results} \label{sec:results}

This section presents the translation of existing bounds on the quantum gravity energy scale $E_{QG,2}$ to constraints on the SME coefficients. An extensive literature review has been conducted to collect data on $E_{QG,2}$ based on the time-of-flight method. To ensure consistency across the different studies, we first homogenized the reported bounds on $E_{QG,2}$ in terms of confidence level, systematic uncertainties, and possible discrepancies in the series expansion, Eq.~\ref{eq:group_velocity_liv}, of the group velocity of the photon.

Constraints on the sum of the non-birefringent coefficients of the photon sector, 
$c^{(6)}_{(I)jm}$, 
are obtained following the formalism introduced in the previous section. Then, we compare the bounds with those reported in the data tables~\citep{Kostelecky:2011}. Afterwards, we present a method to derive bounds on the individual coefficients $c^{(6)}_{(I)jm}$ and list the obtained bounds. 

\subsection{Bounds on the non-birefringent coefficients of the photon sector} \label{subsec:cI_coeff}

A summary of the most stringent lower bounds on $E_{QG,1}$ and $E_{QG,2}$ based on time-of-flight studies 
is given in Table~\ref{tab:LIV_SME_conversion}, which includes both superluminal ($s_\pm=-1$) and subluminal ($s_\pm=+1$) scenarios. All bounds were derived following the cosmological model proposed by \citet{JacobPiran:2008} and expressed at a 95\% confidence level (CL). 
Only results obtained using the maximum likelihood method~\citep{Wilks:1938} have been considered. 


Several bounds presented in Table~\ref{tab:LIV_SME_conversion} required corrections to be comparable. Specifically, those bounds marked with an asterisk ($^*$) had been initially published omitting the pre-factor $(n+1)/2$ in the quadratic term of the photon group velocity expansion (Eq.~\ref{eq:group_velocity_liv}). We incorporated that factor post-hoc to their original publication. In addition, the entries marked with a dagger ($^\dagger$) account for corrections due to systematic uncertainties in the energy scale that were omitted in the original publication. These systematic uncertainties depend on the instrument: a 10\% uncertainty is considered for Fermi-LAT \cite{Vasileiou:2013} and an uncertainty of 12\% is assumed for the Water Cherenkov Detector Array (WCDA) as suggested by the LHAASO experiment \cite{Aharonian_2021}, while a 20\% uncertainty is applied for Imaging Atmospheric Cherenkov Telescopes (IACTs), like MAGIC and H.E.S.S. \cite{magic_tech_part2,hess_tech}. The corresponding bounds on the sum of the SME coefficients $c^{(d)}_{(I)jm}$ were then derived using Eq.~\ref{eq:relation_SME_LIV} and are also listed in Table~\ref{tab:LIV_SME_conversion}. 



\begin{table}[h!]
\centering
\resizebox{\columnwidth}{!}{%
    \begin{tabular}{cccccccccc}
    \toprule \textbf{Source} & \textbf{RA} ($^\circ$) & \textbf{Dec} ($^\circ$)& \textbf{Redshift} & \textbf{Type} & \begin{tabular}{c} $\mathbf{E_{QG,1}}$ \\ ($\times 10^{19}$ GeV) \end{tabular}  & \begin{tabular}{c} $\mathbf{E_{QG,2}}$ \\ ($\times 10^{10}$ GeV) \end{tabular} & \begin{tabular}{c} $\mathbf{\sum_{j m}  Y_{j m} c_{(I) j m}^{(6)}}$ \\ ($\times 10^{-22}$ GeV$^{-2}$) \end{tabular}  & \begin{tabular}{c} $\mathbf{\sum_{j m}  Y_{j m} c_{(I) j m}^{(6)}}$ \\ ($\times 10^{-22}$ GeV$^{-2}$) \\ \textit{Two-sided} \end{tabular} & \begin{tabular}{c} \textit{Likelihood} \\ \textit{availability} \end{tabular} \\
    \midrule \begin{tabular}{c} Crab Pulsar$^{\ddagger}$ \\ \cite{PulsarLIV:2017} \end{tabular} & 83.63 & 22.01 & 2.0 kpc & \begin{tabular}{c} $s_\pm=+1$ \\
    $s_\pm=-1$
    \end{tabular}& \begin{tabular}{c}
    $> 0.055$ \\
    $> 0.045$
    \end{tabular} & \begin{tabular}{c}
    $> 5.9$ \\
    $> 5.3$
    \end{tabular} & \begin{tabular}{c}
    $< 1.4$ \\
    \llap{$>$} $-$1.8
    \end{tabular} & \begin{tabular}{c}
    $< 1.7$ \\
    \llap{$>$} $-$2.1
    \end{tabular} & \cmark \\

    \midrule\midrule \begin{tabular}{c} PKS 2155-304 \\ \cite{PKS2155LIV:2011} \end{tabular}   & 329.72 
    & -30.22
    & 0.12 
    &\begin{tabular}{c}
    $s_\pm=+1$\\
    $s_\pm=-1$
    \end{tabular} & \begin{tabular}{c}
    $> 0.21$ \\
    $> 1.5$ 
    \end{tabular}  & \begin{tabular}{c}
    $> 6.4$ \\
    $> 7.0$
    \end{tabular} & \begin{tabular}{c}
    $> 1.2$ \\
    \llap{$>$} $-$1.0
    \end{tabular} & \begin{tabular}{c}
    $> 1.2$ \\
    \llap{$>$} $-$1.0
    \end{tabular}  & \cmark \\
    \midrule \begin{tabular}{c} PG 1553+113$^{\ddagger}$ \\ \cite{Abramowski_2015,DorigoJones:2022} \end{tabular} & 238.94 
    & 11.19
    & 0.43 
    &\begin{tabular}{c}
    $s_\pm=+1$\\
    $s_\pm=-1$
    \end{tabular} & \begin{tabular}{c}
    $> 0.041$ \\
    $> 0.028$
    \end{tabular} & \begin{tabular}{c}
    $> 2.1$ \\
    $> 1.7$
    \end{tabular} & \begin{tabular}{c}
    $< 11$ \\
    \llap{$>$} $-$18
    \end{tabular} & \begin{tabular}{c}
    $< 14$ \\
    \llap{$>$} $-$21
    \end{tabular} & \xmark \\
    \midrule \begin{tabular}{c} Mrk 501$^{*,\ddagger}$ \\ \cite{Abdalla:2019_LIV} \end{tabular} & 253.47 
    & 39.76
    & 0.034 & \begin{tabular}{c}
    $s_\pm=+1$\\
    $s_\pm=-1$
    \end{tabular} & \begin{tabular}{c}
    $> 0.036$ \\
    $> 0.026$
    \end{tabular} & \begin{tabular}{c}
    $> 10.4$ \\
    $> 8.9$
    \end{tabular} & \begin{tabular}{c}
    $< 0.46$ \\
    \llap{$>$} $-$0.63
    \end{tabular} & \begin{tabular}{c}
    $< 0.57$ \\
    \llap{$>$} $-$0.73
    \end{tabular} & \cmark \\
    \midrule \begin{tabular}{c} Mrk 421 \\ \cite{Mrk421LIV:2024} \end{tabular} & 166.08 
    & 38.19
    & 0.031 & \begin{tabular}{c}
    $s_\pm=+1$\\
    $s_\pm=-1$
    \end{tabular} & \begin{tabular}{c}
    $> 0.036$ \\
    $> 0.027$
    \end{tabular} & \begin{tabular}{c}
    $> 2.5$ \\
    $> 2.6$
    \end{tabular} & \begin{tabular}{c}
    $< 8.0$ \\
    \llap{$>$} $-$7.4
    \end{tabular} & \begin{tabular}{c}
    $< 8.0$ \\
    \llap{$>$} $-$7.4
    \end{tabular} & \cmark \\
    \midrule\midrule \begin{tabular}{c} GRB 090510$^{\dagger,\ddagger}$ \\ \cite{Vasileiou:2013} \end{tabular} & 333.55
    & -26.58
    & 0.90 
    & \begin{tabular}{c}
    $s_\pm=+1$\\
    $s_\pm=-1$
    \end{tabular} & \begin{tabular}{c}
    $> 4.7$ \\
    $> 10$
    \end{tabular} & \begin{tabular}{c}
    $> 7.7$ \\
    $> 8.5$
    \end{tabular} & \begin{tabular}{c}
    $< 0.84$ \\
    \llap{$>$} $-$0.70
    \end{tabular} & \begin{tabular}{c}
    $< 0.98$ \\
    \llap{$>$} $-$0.85
    \end{tabular} & \cmark (?) \\
    \midrule \begin{tabular}{c} GRB 080916C$^{\dagger,\ddagger}$ \\ \cite{Vasileiou:2013} \end{tabular} & 119.85
    & -56.64 
    &4.4 
    & \begin{tabular}{c}
    $s_\pm=+1$\\
    $s_\pm=-1$
    \end{tabular} & \begin{tabular}{c}
    $> 0.20$ \\
    $> 0.18$
    \end{tabular} & \begin{tabular}{c}
    $> 0.32$ \\
    $> 0.31$
    \end{tabular} & \begin{tabular}{c}
    $<$ 5.0$\times$10$^2$ \\
    \llap{$>$} $-$5.3$\times$10$^2$
    \end{tabular} & \begin{tabular}{c}
    $<$ 6.0$\times$10$^2$ \\
    \llap{$>$} $-$6.3$\times$10$^2$
    \end{tabular} & \cmark (?) \\
    \midrule \begin{tabular}{c} GRB 090902B$^{\dagger,\ddagger}$ \\ \cite{Vasileiou:2013} \end{tabular} & 264.94 
    & 27.32
    & 1.8 %
    & \begin{tabular}{c}
    $s_\pm=+1$\\
    $s_\pm=-1$
    \end{tabular}& \begin{tabular}{c}
    $> 0.11$ \\
    $> 0.33$
    \end{tabular} & \begin{tabular}{c}
    $>$ 0.58 \\
    $>$ 0.58
    \end{tabular} & \begin{tabular}{c}
    $<$ 1.5$\times$10$^2$ \\
    \llap{$>$} $-$1.5$\times$10$^2$
    \end{tabular} & \begin{tabular}{c}
    $<$ 1.8$\times$10$^2$ \\
    \llap{$>$} $-$1.8$\times$10$^2$
    \end{tabular} & \cmark (?) \\
    \midrule \begin{tabular}{c} GRB 090926A$^{\dagger,\ddagger}$ \\ \cite{Vasileiou:2013} \end{tabular} & 353.40
    & -66.32
    & 2.1 
    & \begin{tabular}{c}
    $s_\pm=+1$\\
    $s_\pm=-1$
    \end{tabular}& \begin{tabular}{c}
    $> 1.1$ \\
    $> 0.15$
    \end{tabular} & \begin{tabular}{c}
    $> 0.43$ \\
    $> 0.28$
    \end{tabular} & \begin{tabular}{c}
    $<$ 2.7$\times$10$^2$ \\
    \llap{$>$} $-$6.4$\times$10$^2$
    \end{tabular} & \begin{tabular}{c}
    $<$ 3.6$\times$10$^2$ \\
    \llap{$>$} $-$7.3$\times$10$^2$
    \end{tabular} & \cmark (?) \\
    \midrule \begin{tabular}{c} GRB 190114C$^{\dagger}$ \\ \cite{GRB1901114LIV:2020} \end{tabular} & 54.51 
    & -26.95 
    & 0.42
    & \begin{tabular}{c}
    $s_\pm=+1$\\
    $s_\pm=-1$
    \end{tabular}& \begin{tabular}{c}
    $> 0.46$ \\
    $> 0.44$
    \end{tabular} & \begin{tabular}{c}
    $> 5.0$ \\
    $> 4.5$
    \end{tabular} & \begin{tabular}{c}
    $< 2.0$ \\
    \llap{$>$} $-$2.5
    \end{tabular} & \begin{tabular}{c}
    $< 2.0$ \\
    \llap{$>$} $-$2.5
    \end{tabular} & \cmark \\
    \midrule \begin{tabular}{c} GRB 221009A$^{\dagger}$ \\ \cite{lhaaso_2024} \end{tabular} & 288.26
    & 19.77
    & 0.15 
    & \begin{tabular}{c}
    $s_\pm=+1$\\
    $s_\pm=-1$
    \end{tabular}& \begin{tabular}{c}
    $> 8.0$ \\
    $> 8.8$
    \end{tabular} & \begin{tabular}{c}
    $> 55$ \\
    $> 56$
    \end{tabular} & \begin{tabular}{c}
    $< 0.014$ \\
    \llap{$>$} $-$0.013
    \end{tabular} & \begin{tabular}{c}
    $< 0.014$ \\
    \llap{$>$} $-$0.013
    \end{tabular} & \cmark (?) \\
    \bottomrule
    \end{tabular}
}
\caption{List of the most stringent lower bounds on $E_{QG,1}$ and $E_{QG,2}$ based on time-of-flight studies. The table is divided by source type: pulsars (first row), AGNs (second to fifth rows) and GRBs (sixth to last rows). $s_\pm=-1$ and $s_\pm=+1$ represent superluminal and subluminal behaviours, respectively. All lower bounds are expressed at $95 \%$ CL. The $^*$ refers to the correction due to the missing term $(n+1)/2$ for the photon group velocity, and $^\dagger$ indicates the correction to the energy scale due to systematics. A 10\% systematic uncertainty due to instrumental effects is used for Fermi-LAT \cite{Vasileiou:2013}) and a 12\% is considered for LHAASO~\cite{lhaaso_2024}, while 20\% is applied to IACTs (\cite{magic_tech_part2,hess_tech}). The $^\ddagger$ marks those analyses originally carried out using a one-sided Gaussian distribution, which were converted into equivalent two-sided Gaussian bounds. Those two-sided, non-birefringent SME coefficients are then shown in the penultimate column. The final column indicates whether the detailed likelihood profile for each source is publicly available: \cmark~denotes complete availability, \cmark~(?) indicates partial or incomplete availability, and \xmark~indicates it is unavailable. }\label{tab:LIV_SME_conversion}
\end{table}

Comparing our results
with those presented in the 
data tables~\citep{Kostelecky:2011}, we observe the following differences: 
First, this work includes several new astrophysical sources not formerly covered in the data tables, specifically the Crab Pulsar~\cite{PulsarLIV:2017}, the AGNs PG~1553+113~\cite{Abramowski_2015} and Mrk~421~\cite{Mrk421LIV:2024}, and the gamma-ray bursts GRB~190114C~\cite{GRB1901114LIV:2020} and GRB~221009A~\cite{lhaaso_2024}. Among these, 
the results on Mrk~421 and GRB~221009A have been published only recently. We carefully revised the corresponding bounds on the quantum gravity energy scale $E_{QG,n}$ from these sources and converted both superluminal and subluminal on $E_{QG,2}$ bounds to non-birefringence SME parameters using Eq.~\ref{eq:relation_SME_LIV}. Notably, the original studies of GRB~190114C~\cite{GRB1901114LIV:2020} and GRB~221009A~\cite{lhaaso_2024} did not account for systematic uncertainties in the energy scale; we have incorporated these corrections by applying 
20\% and 12\% uncertainty, respectively. 

It is also important to highlight the inclusion of a pulsar ---namely, the Crab Pulsar --- for the first time in such an analysis.
Due to their high stability and predictable emission profiles, pulsars serve as particularly valuable benchmarks for time-of-flight LIV studies. 
Their inclusion broadens the scope of this work, increasing the diversity of source types and enhancing the overall robustness of the analysis.   
Moreover, we have updated 
the 
bounds from Mrk~501, originally presented in the data tables based on MAGIC observations~\cite{magic_mrk501}, 
to reflect the more recent and stringent results from the H.E.S.S. collaboration~\cite{Abdalla:2019_LIV}, who reported a stronger flare in June 2014 with an extended energy range up to 20~TeV.  A correction for the pre-factor $(n+1)/2$ in the quadratic term of the photon group velocity was required and implemented accordingly. 
Similarly, 
for PKS~2155-304~\cite{PKS2155LIV:2011}, which is found 
in the data tables, we applied the same correction for the pre-factor $(n+1)/2$, 
yielding a tighter bound on $E_{QG,2}$ and consequently on 
the non-birefringent coefficients of the photon sector.

For the bounds obtained from 
GRB~090510~\cite{Vasileiou:2013}, GRB~080916C~\cite{Vasileiou:2013}, GRB~090902B~\cite{Vasileiou:2013} and GRB~090926A~\cite{Vasileiou:2013}, we applied a correction corresponding to 10\% systematic uncertainty on the energy scale of Fermi-LAT. We also note that only results obtained via the maximum likelihood method from \citet{Vasileiou:2013} were selected and translated into bounds on the SME coefficients. 
When attempting to reproduce the bounds on SME coefficients reported by \citet{Vasileiou:2013} -- which, to our knowledge, is the only publication in the literature directly converting time-of-flight results to the SME framework -- 
we found several discrepancies. A possible missing factor of $(d-3)$ may explain the mismatch only partially. 
In summary, the corrected 
bounds from GRB~090510, GRB~080916C, GRB~090902B, and GRB~090926A lead to slightly tighter constraints on the SME coefficients compared to \citet{Vasileiou:2013}. 

Several time-of-flight studies 
 used one-sided Gaussian confidence intervals (CI). We have converted these into their  equivalent two-sided Gaussian CIs to ensure consistency across all bounds. 
Analyses that originally employed a one-sided Gaussian distribution are denoted 
with a double-dagger ($^\ddagger$) in the table.  These include the Crab Pulsar~\cite{PulsarLIV:2017}, PG~1553+113~\cite{Abramowski_2015}, Mrk~501~\cite{Abdalla:2019_LIV}, GRB~090510~\cite{Vasileiou:2013}, GRB~080916C~\cite{Vasileiou:2013}, GRB~090902B~\cite{Vasileiou:2013} and GRB~090926A~\cite{Vasileiou:2013}. 
To perform the conversion, we reviewed the profile likelihoods associated with each source and marked their availability in the final column of Table~\ref{tab:LIV_SME_conversion}. Whereas the majority of the studies report the profile likelihood -- with the exception of PG~1553+113~\cite{Abramowski_2015} -- 
several publications do not provide the best-fit values that minimise the profile likelihood, making it difficult to perform an accurate two-sided conversion, especially when these profile likelihoods are not well-behaved. 
In particular,  \citet{Vasileiou:2013} and \citet{lhaaso_2024} did not publish the calibrated likelihoods, i.e., the reference likelihood functions from which the subsequent bound on $E_{QG,2}$ has been derived. Therefore, to ensure coherence across all datasets, we assumed symmetric and well-behaved profile likelihoods, using the central values of the reported bounds as proxies for the best-fit values. Under this assumption, we have converted the one-sided 95\%~CL bounds into their equivalent for a two-sided 95\%~CL. 

In summary, new and updated bounds on non-birefringent SME coefficients of the photon sector derived from time-of-flight studies have been presented in this work. The corrected bounds yield slightly more stringent constraints on LIV effects compared to those in the data tables compilation~\cite{Kostelecky:2011}. While GRB~221009A~\cite{lhaaso_2024} stands out for providing the most stringent bound to date, it remains an isolated event. 



\subsection{Bounds on individual non-birefringent SME coefficients}
\label{subsec:kI_coeff}

Any time delay measurement from photons arriving from a certain source in the sky observed from direction $\hat{n}_i$ leads, in some form, to a posterior probability distribution for the weighted sum of SME coefficients $\theta_i := \sum_{j,m}{}_0Y_{jm}(\hat{n}_i)c_{(I)jm}^{(d)}$
\begin{equation}
P_i\left(\theta_i;\vec{\nu_i}\,\big|\,D_i\right) \qquad  i = 1,\ldots,s \quad,
\label{complex-sum}
\end{equation}
in which $D_i$ denotes the $i^\mathrm{th}$ of a total $s$ time delay measurements and $\vec{\nu}_i$ a possible set of nuisance parameters for $D_i$, for instance, unknowns in the instrumental acceptance or intrinsic source-dependent effects. 
If the nuisance parameters have been marginalized or a profile likelihood~\cite{MurphyVanderWaart:2000} calculated, a probability distribution depending only on data can be calculated. In the case of large samples, the likelihood $\mathcal{L}_i(D_i|\theta_i)$ will approach a normal distribution~\cite{Wilks:1938} for one degree of freedom, 
with a given central mode $\theta_{m,i}$ (which maximizes $\mathcal{L}_i$) and width $\sigma_{\theta_i}$. In such a case, $\theta_{m,i}$ will be found close to the expectation value $E[\theta_i]$ and $\sigma_{\theta_i}^2$ close to its variance $s^2[\theta_i]$. Unfortunately, very often~\citep{Vasileiou:2013,PulsarLIV:2017,Wei:2022}, statistics of those events with the highest energies -- which receive a strong weight in the likelihood in the case of $d>5$ (see Eq.~\eqref{eq:time_delay_SME}) -- is low, also due to the typical power-law spectra of the involved sources. Together with asymmetric systematics $\vec{\nu}_i$,  this leads to not ``well-behaved" profile likelihoods, which show serious skewness and tails exceeding the Gaussian expectation. In the best cases, a calibration of $\mathcal{L}_i$ is carried out~\citep{PKS2155LIV:2011,Vasileiou:2013,PulsarLIV:2017}, e.g., through simulations or bootstrapping methods, to match the coverage properties of the actual likelihood with those expected from a normal distribution\footnote{Although unfortunately, the likelihood calibration recipes are often only applied to derive the corresponding bounds, without publishing the recipe itself (see Section~\ref{sec:results} and Table~\ref{tab:LIV_SME_conversion}).}. Authors then publish (asymmetric) confidence intervals around $\theta_{m,i}$, or only just the 95\%~CL bounds, assuming that a null-measurement was the most plausible outcome of the measurement. 
The posterior probability distribution $P_i$ and the likelihood $\mathcal{L}_i$ are related through Bayes theorem and may incorporate prior knowledge on the coefficients. In the absence of prior knowledge about the coefficient, a flat prior (as used, for instance, in \citet{Wei:2022}) is not recommended~\citep{Lista:2016,Barlow:2021} because of its lack of scale and parameter transformation invariance. Better priors have been proposed~\citep{Jeffreys:1946,Demortier:2010}, but are not always used. 

Although it is standard to combine likelihoods (sum $\ln(\mathcal{L}_i)$ from different measurements of one parameter to obtain a combined likelihood from all measurements, the inverse is not possible to our knowledge, that is, to combine likelihoods of one or various parameters $\theta_i$ that are by themselves linear combinations of unknowns. The latter is the case for the coefficients $c_{(I)jm}^{(d)}$, on which likelihoods have been derived only for their linear combination $\sum_{j,m}{}_0Y_{jm}(\hat{n}_i)c_{(I)jm}^{(d)}$. 
The spherical harmonics ${}_0Y_{jm}(\hat{n_i})$ and the coefficients $c_{(I)jm}^{(d)}$ are moreover complex and satisfy the relations
\begin{align}
{}_0Y_{jm}(\hat{n}_i) &= (-1)^m {}_0Y_{j(-m)}(\hat{n}_i)^* \\
c_{(I)jm}^{(d)} &= (-1)^m c_{(I)j(-m)}^{(d)*}~.
\end{align}

Therefore, $c_{(I)j0}^{(d)}$ is real, while for $m \ne 0$ the pair of complex coefficients $c_{(I)jm}^{(d)}$ and $c_{(I)j(-m)}^{(d)}$ are co-dependent but represent two real degrees of freedom. For fixed (even) $d$, the number of real degrees of freedom for the set of coefficients $c_{(I)jm}^{(d)}$ is 
therefore
$N=(d-1)^2$.
We can show them explicitly by expressing the left-hand side of Eq.~\eqref{complex-sum} as
\begin{equation}
\sum_{j=0}^{d-2}\left({}_0Y_{j0}(\hat{n}_i)c_{(I)j0}^{(d)}
+ 2\sum_{m=1}^j\Bigl(\bigl(\re{{}_0Y_{jm}(\hat{n}_i)}\bigr)\bigl(\re{c_{(I)jm}^{(d)}}\bigr)
- \bigl(\im{{}_0Y_{jm}(\hat{n}_i)}\bigr)\bigl(\im{c_{(I)jm}^{(d)}}\bigr)\Bigr)\right).
\label{real-sum}
\end{equation}
%

Identifying the coefficients $c_{(I)j0}^{(d)}$, $\re c_{(I)jm}^{(d)}$ and $\im c_{(I)jm}^{(d)}$ ($m>0$) with the (real) $N$-dimensional vector $x_j$,  we calculate a global probability distribution function for the set of independent coefficients 
\begin{equation}
P(x_1,\ldots,x_N) = \prod_{i=1}^{s} P_i(x_1,\ldots,x_N|D_i)~.
\label{eq:Px1toxN}
\end{equation}

We assume a normal distribution as best guess for $P_i(x_1,\ldots,x_N|D_i)$, leading to 
%
%
%
\begin{equation}
P(x_1,\ldots,x_N) \propto \exp\left[-\sum_{i=1}^s\frac{\left(\sum_{j=1}^Na_{ij}x_j - \mu_i\right)^2}{2\sigma_i^2}\right] \ \,
\label{eq:Pxexponent}
\end{equation}
with $a_{ij}$ being the values of the spherical harmonics in the direction of the source. 
Now, as outlined in \citet{Dagostini:2003,DAgostini:2004}, $\theta_{m,i}$ would be a wrong approximation for $\mu_i$ in the case of skewed likelihoods. We adopt, therefore, the prescriptions of \citet{DAgostini:2004} to approximate instead $E[\theta_i]$ and $s^2(\theta_i)$ from the asymmetric bounds provided and (if available) the central value $\theta_{m,i}$. 
Whenever publications become available that include the full information on the calibrated likelihoods, including expectation values and variances, this part of our analysis can be improved in that respect.

For convenience, redefining $\tilde{a}_{ij} = a_{ij}/\sigma_i$ and $\tilde{g}_i = \mu_i/\sigma_i$, we have 
\begin{align}
P(x_1,\ldots,x_N) &\propto \exp\left[-\frac{1}{2}\sum_{i=1}^s\left(\sum_{j=1}^N \tilde{a}_{ij}x_j - \tilde{g_i}\right)^2\right]\nonumber\\
&= \exp\left[-\tfrac{1}{2}x^T(\tilde{a}^T\tilde{a})x + (\tilde{g}^T\tilde{a})x - \tfrac{1}{2}\tilde{g}^T\tilde{g}\right] \nonumber\\
&= \exp\left[-\tfrac{1}{2}\left( (x-\langle x \rangle)^T A (x-\langle x \rangle) \right)  \right]  
\label{Px-matrix}
\end{align}
 where the last two lines are expressed using matrix notation.  We see that $P(x_1,\ldots,x_N)$ amounts to a multidimensional Gaussian distribution that is controlled by the $N\times N$ symmetric matrix $\tilde{a}^T\tilde{a}$ and the (co-)vector $\tilde{g}^T\tilde{a}$.
 In the last line of Eq.~\eqref{Px-matrix}, we  reformulate $P(x_1,\ldots,x_N)$ in a more convenient form  
with $A=\tilde{a}^T\tilde{a}$ and $\langle x \rangle=A^{-1} \tilde{a}^T \tilde{g}$. 

The easiest way to understand the form of the probability distribution \eqref{Px-matrix} is by diagonalizing the matrix $\tilde{a}^T\tilde{a}$. Note that this can always be done for a symmetric matrix by applying an orthogonal transformation $Q$:
$x \to y = Qx$, $\tilde{a}^T\tilde{a} \to M = Q \tilde{a}^T\tilde{a} Q^{-1} = Q \tilde{a}^T\tilde{a} Q^T$ ($Q^{-1} = Q^T$ for orthogonal~$Q$), so that\footnote{Note that \textit{mpmath}'s \textit{eighe} function returns $Q^T$, aside of $M$, instead of $Q$.}
%
\begin{equation}
M_{ij} = \begin{pmatrix}
M_{11} & 0 & 0 & \cdots & 0\\
0 & M_{22} & 0 & \cdots & 0\\
\vdots & \vdots & & & \vdots\\
0 & 0 & 0 & \cdots & M_{nn}
\end{pmatrix} ~.
\end{equation}
The eigenvalues $M_{ii}$ are guaranteed to be nonnegative\footnote{If $(\tilde{a}^T\tilde{a})x = \lambda x$ for nonzero eigenvector $x$, it follows that $\lambda (x^Tx) = x^T(\tilde{a}^T\tilde{a})x = (\tilde{a}x)^T(\tilde{a}x)$.  As $x^Tx > 0$ and $(\tilde{a}x)^T(\tilde{a}x)\ge 0$ (both expressions are equal to a sum of squares), it follows that $\lambda \ge 0$.}. The rank of the matrix $M$, $\text{rank}(M) = r$, is equal to the number of positive eigenvalues, which can at most be equal to the minimum of the values of $N$ and $s$.  Since the number of measurements $s \ge N$, we can make $r = N$. 

It is convenient to arrange the eigenvalues so that $M_{ii} > 0$ for $i=1,\ldots,r$,
while $M_{jj} = 0$ for $j = r+1,\ldots,N$. We have
\begin{equation}
P(y) \propto \exp\left[-\tfrac{1}{2}(y - \langle y\rangle)^T M(y - \langle y\rangle)\right]~. 
\label{eq:proby}
\end{equation}
This distribution is a direct product of Gaussians for the components $y_i$, $i=1,\ldots,r$, such that the variance for the coordinate $y_i$ is equal to $M_{ii}^{-1}$.
Moreover, it is easy to see that the distribution yields an average value for $y_i$ equal to
\begin{equation}
\langle y_i \rangle = \frac{\bigl(Q\tilde{a}^T\tilde{g}\bigr)_i}{M_{ii}} \qquad i=1,\ldots,r~
\label{expec-y}.
\end{equation}

%
We can obtain the distribution in terms of the components of the vector $x$ by applying the orthogonal transformation (i.e., multidimensional rotation) $Q^T$. It can be visualized as a multidimensional ellipsoid, centered around the expectation value of the vector $x$,
\begin{equation}
\langle x\rangle = Q^T\langle y \rangle
\end{equation}
where the components of the vector $\langle y \rangle$ are given by Eq.~\eqref{expec-y}. Eq.~\eqref{eq:Qij} provides the matrix $Q$ which leads to rotated variables with a precision of better than 0.5\%. 

A reduced distribution function for any of the components $x_i$ of the vector $x$ can be obtained by integrating over (marginalizing) all the remaining components, for instance:
\begin{equation}
\tilde{P}(x_1) = \int dx_2\ldots dx_n\,P(x_1,\ldots,x_n).
\end{equation}
This integral is well defined if $r = N$, in which case it leads to
\begin{equation}
\tilde{P}(x_1) \propto \exp\left[-\tfrac12 \bigl(x_1 - \langle x_1\rangle\bigr)^2/\bigl((\tilde{a}^T\tilde{a})^{-1}\bigr)_{11}\right]
\end{equation}
which is a Gaussian distribution with variance $s^2(x_1)=\bigl((\tilde{a}^T\tilde{a})^{-1}\bigr)_{11}$ \footnote{Note that $\bigl((\tilde{a}^T\tilde{a})^{-1}\bigr)_{11} \ne \bigl((\tilde{a}^T\tilde{a})_{11}\bigr)^{-1}$ .}.

A similar reasoning using the properties of the expectation value: $E[\sum_i a_i x_i] = \sum_i a_i E[x_i]$, $\sigma^2[\sum_i a_i x_i] = \sum_i a_i^2 \sigma^2[x_i]$ leads to the same solutions for $\langle x_i \rangle$ and $s^2(x_i)$. Note that  \citet{Kislat:2015} used only the second equation, on 95\% asymmetric bounds $\sigma^2[\sum_i a_i x_i]$, without specifying how the asymmetry was treated, and without taking into account the expectation values $E[\sum_i a_i x_i]$.

If $r < N$ 
, it is not possible to obtain reduced distribution functions for the individual components $x_i$. In that case, the best one can do is to determine the eigenvectors in $x$ space corresponding to the nonzero eigenvalues $M_{11}, \ldots, M_{rr}$. They are given by the first $r$ columns of the matrix $Q^T$ (or, equivalently, the first $r$ rows of the matrix $Q$). We can then express the probability distribution in the form
\begin{equation}
P(x_1,\ldots,x_n) = \exp\left[-\frac12 \sum_{i=1}^r M_{ii}\left(\sum_{j=1}^N Q_{ij}(x_j - \langle x_j\rangle)\right)^2\right]
\end{equation}
which shows explicitly that the linear combination $\sum_{j=1}^N Q_{ij}x_j$, for $i=1,\ldots, r$, is bounded with variance $(M_{ii})^{-1}$.

We have solved both Eqs.~\eqref{Px-matrix} and~\eqref{eq:proby} for their expectation values and variances, shown in Tables~\ref{tab:coeffsA} and~\ref{tab:coeffsB}. For the inversions, we adopted the estimated expectation values for $\sum_{j,m}{}_0Y_{jm}(\hat{n}_i)c_{(I)jm}^{(d)}$ of Table~\ref{tab:LIV_SME_conversion} for $\mu_i$ ($i=1,\ldots,11$), following $E[x_i] \approx 1/2\cdot(\textit{UL}_i+\textit{LL}_i)$, where \textit{UL} and \textit{LL} correspond to $s_\pm=+1$ and $s_\pm=-1$ of the column \textit{Two-sided} 
of Table~\ref{tab:LIV_SME_conversion},
and $\sigma_i\approx 1/4\cdot(\textit{UL}_i+\textit{LL}_i)$.  For the remaining bounds, needed to fill $r$ up to the minimum of $N\geq 25$ measurements, we have adopted the 24 measurements of Table~III of~\citet{Kislat:2015}, the bounds of Table~1 of \citet{Wei:2017} (corrected for their fit $\chi^2/\mathrm{ndf}$), 29 bounds on GRB021206B from \citet{Boggs:2004,Kostelecky:2008be} and the bounds from GRBs of Table~2 of~\citet{Wei:2022}, following the prescription of \citet{DAgostini:2004}: $E[x_i] \approx \theta_{m,i}+1/2\cdot(\Delta_{+,i}+\Delta_{-,i})$ and  $\sigma_i \approx 1/4\cdot(\Delta_{+,i}+\Delta_{-,i})$, where $\Delta_{+,i}$ and $\Delta_{-,i}$ represent the asymmetric 95\%~CL uncertainties with respect to the mode $\theta_{m,i}$. With these additions, a total of 65 datasets are available. The approximated expectation values and standard deviations of the 54 added data sets for this part of the analysis are shown in Tables~\ref{tab:LIV_table_AGNs} and~\ref{tab:LIV_table_GRBs}.

On the technical side of the implementation, we observed that an internal precision of at least 23 decimal places\footnote{Available, e.g., in \textit{Matlab} or the \textit{mpmath} library in \textit{Python}.} is required for stable solutions. 

We observed that the direct bounds on $c_{(I)jm}^{(d)}$ do slightly improve as more and more of the less constraining bounds $N>25$ are added, although they also become more asymmetric, due to the inherent asymmetry of many of the GRB measurements of~\citet{Wei:2022}. Table~\ref{tab:coeffsA} shows the solution for all 65 measurements used. If only 25 are taken, the bounds worsen by about 10\%. On the contrary, the bounds on the rotated parameters $y_i$ of Table~\ref{tab:coeffsB} remain almost unaffected by the addition of further measurements beyond $N=25$, except for the very last parameter $y_{25}$, as expected.

\begin{table}[h!]
\centering
\begin{tabular}{ccccc}
\toprule\addlinespace[0.5ex] 
\textbf{Coefficient} & \begin{tabular}{c} $\mathbf{\langle x_i\rangle}$ \\ ($\times 10^{-15}~\mathrm{GeV}^{-2}$) \end{tabular} & \begin{tabular}{c} $\mathbf{\sqrt{A_{ii}^{-1}}}$ \\ ($\times 10^{-15}~\mathrm{GeV}^{-2}$) \end{tabular} & \begin{tabular}{c} \textbf{LL (95\% CL)} \\ ($\times 10^{-15}~\mathrm{GeV}^{-2}$) \end{tabular} & 
\begin{tabular}{c} \textbf{UL (95\% CL)} \\ ($\times 10^{-15}~\mathrm{GeV}^{-2}$) \end{tabular} 
\\ \midrule
$c_{(I)00}^{(6)}$      &  \llap{$-$}0.5  &  1.5  &  \llap{$-$}3.4  &  2.4 \\
$c_{(I)10}^{(6)}$      &  0.0  &  0.8  &  \llap{$-$}1.5  &  1.6 \\
$\re{c_{(I)11}^{(6)}}$ &  0.1  &  0.7  &  \llap{$-$}1.3  &  1.4 \\
$\im{c_{(I)11}^{(6)}}$ &  \llap{$-$}0.3  &  1.1  &  \llap{$-$}2.5  &  1.9 \\
$c_{(I)20}^{(6)}$      &  0.2  &  1.1  &  \llap{$-$}2.0  &  2.4 \\
$\re{c_{(I)21}^{(6)}}$ &  0.0  &  0.5  &  \llap{$-$}1.0  &  1.0 \\
$\im{c_{(I)21}^{(6)}}$ &  \llap{$-$}0.2  &  0.4  &  \llap{$-$}1.0  &  0.7 \\
$\re{c_{(I)22}^{(6)}}$ &  0.1  &  0.5  &  \llap{$-$}1.0  &  1.2 \\
$\im{c_{(I)22}^{(6)}}$ &  0.2  &  1.0  &  \llap{$-$}1.7  &  2.1 \\
$c_{(I)30}^{(6)}$      &  \llap{$-$}0.2  &  0.8  &  \llap{$-$}1.8  &  1.4 \\
$\re{c_{(I)31}^{(6)}}$ &  \llap{$-$}0.4  &  0.9  &  \llap{$-$}2.2  &  1.3 \\
$\im{c_{(I)31}^{(6)}}$ &  0.4  &  0.7  &  \llap{$-$}1.1  &  1.8 \\
$\re{c_{(I)32}^{(6)}}$ &  \llap{$-$}0.3  &  0.3  &  \llap{$-$}0.9  &  0.3 \\
$\im{c_{(I)32}^{(6)}}$ &  0.5  &  0.5  &  \llap{$-$}0.5  &  1.5 \\
$\re{c_{(I)33}^{(6)}}$ &  \llap{$-$}0.1  &  0.9  &  \llap{$-$}1.9  &  1.8 \\
$\im{c_{(I)33}^{(6)}}$ &  \llap{$-$}0.1  &  0.7  &  \llap{$-$}1.6  &  1.3 \\
$c_{(I)40}^{(6)}$      &  \llap{$-$}0.7  &  0.7  &  \llap{$-$}2.0  &  0.6 \\
$\re{c_{(I)41}^{(6)}}$ &  \llap{$-$}0.2  &  0.2  &  \llap{$-$}0.7  &  0.2 \\
$\im{c_{(I)41}^{(6)}}$ &  \llap{$-$}0.3  &  0.4  &  \llap{$-$}1.0  &  0.4 \\
$\re{c_{(I)42}^{(6)}}$ &  0.1  &  0.4  &  \llap{$-$}0.7  &  1.0 \\
$\im{c_{(I)42}^{(6)}}$ &  \llap{$-$}0.0  &  0.5  &  \llap{$-$}1.0  &  1.0 \\
$\re{c_{(I)43}^{(6)}}$ &  \llap{$-$}0.2  &  0.2  &  \llap{$-$}0.6  &  0.2 \\
$\im{c_{(I)43}^{(6)}}$ &  \llap{$-$}0.2  &  0.2  &  \llap{$-$}0.6  &  0.1 \\
$\re{c_{(I)44}^{(6)}}$ &  0.1  &  0.3  &  \llap{$-$}0.5  &  0.7 \\
$\im{c_{(I)44}^{(6)}}$ &  0.0  &  0.4  &  \llap{$-$}0.9  &  0.9 \\
\bottomrule
\end{tabular}
\caption{\label{tab:coeffsA} Expectation values (second column), standard deviations (third column), 95\% CL lower (fourth column) and upper bounds (last column) for the 25 non-birefringent coefficients of the SME photon sector for $d=6$. }
\end{table}

\begin{table}[h!]
\centering
\begin{tabular}{ccccc} 
\toprule\addlinespace[0.5ex] 
\textbf{Coefficient} & \begin{tabular}{c} $\mathbf{\langle y_i\rangle}$ \\ ($\times 10^{-22}~\mathrm{GeV}^{-2}$) \end{tabular} & \begin{tabular}{c} $\mathbf{\sqrt{M_{ii}^{-1}}}$ \\ ($\times 10^{-22}~\mathrm{GeV}^{-2}$) \end{tabular} & \begin{tabular}{c} \textbf{LL (95\% CL)} \\ ($\times 10^{-22}~\mathrm{GeV}^{-2}$) \end{tabular} & 
\begin{tabular}{c} \textbf{UL (95\% CL)} \\ ($\times 10^{-22}~\mathrm{GeV}^{-2}$) \end{tabular} 
\\ \midrule
$y_{1}$ &\llap{$-$}9.6$\times$10$^{-5}$  &  3.5$\times$10$^{-3}$  & \llap{$-$}7.0$\times$10$^{-3}$  &  6.8$\times$10$^{-3}$ \\
$y_{2}$ & 1.8$\times$10$^{-2}$  &  1.7$\times$10$^{-1}$  & \llap{$-$}3.2$\times$10$^{-1}$  &  3.5$\times$10$^{-1}$ \\
$y_{3}$ & 6.7$\times$10$^{-2}$  &  1.9$\times$10$^{-1}$  & \llap{$-$}3.2$\times$10$^{-1}$  &  4.5$\times$10$^{-1}$ \\
$y_{4}$ & 4.4$\times$10$^{-2}$  &  4.9$\times$10$^{-1}$  & \llap{$-$}9.3$\times$10$^{-1}$  &  1.0$\times$10$^{0}$ \\
$y_{5}$ &\llap{$-$}1.6$\times$10$^{-1}$  &  6.2$\times$10$^{-1}$  & \llap{$-$}1.4$\times$10$^{0}$  &  1.1$\times$10$^{0}$ \\
$y_{6}$ & 3.9$\times$10$^{-2}$  &  1.8$\times$10$^{0}$  & \llap{$-$}3.5$\times$10$^{0}$  &  3.6$\times$10$^{0}$ \\
$y_{7}$ & 1.7$\times$10$^{-1}$  &  2.2$\times$10$^{0}$  & \llap{$-$}4.2$\times$10$^{0}$  &  4.5$\times$10$^{0}$ \\
$y_{8}$ & 1.8$\times$10$^{0}$  &  5.0$\times$10$^{0}$  & \llap{$-$}8.1$\times$10$^{0}$  &  1.2$\times$10$^{1}$ \\
$y_{9}$ &\llap{$-$}1.7$\times$10$^{1}$  &  1.3$\times$10$^{2}$  & \llap{$-$}2.8$\times$10$^{2}$  &  2.5$\times$10$^{2}$ \\
$y_{10}$ & 6.6$\times$10$^{1}$  &  1.5$\times$10$^{2}$  & \llap{$-$}2.3$\times$10$^{2}$  &  3.7$\times$10$^{2}$ \\
$y_{11}$ &\llap{$-$}1.0$\times$10$^{2}$  &  2.0$\times$10$^{2}$  & \llap{$-$}5.0$\times$10$^{2}$  &  3.0$\times$10$^{2}$ \\
$y_{12}$ &\llap{$-$}1.3$\times$10$^{3}$  &  2.5$\times$10$^{3}$  & \llap{$-$}6.4$\times$10$^{3}$  &  3.7$\times$10$^{3}$ \\
$y_{13}$ &\llap{$-$}1.9$\times$10$^{4}$  &  3.7$\times$10$^{4}$  & \llap{$-$}9.3$\times$10$^{4}$  &  5.6$\times$10$^{4}$ \\
$y_{14}$ & 7.0$\times$10$^{3}$  &  6.5$\times$10$^{4}$  & \llap{$-$}1.2$\times$10$^{5}$  &  1.4$\times$10$^{5}$ \\
$y_{15}$ &\llap{$-$}1.5$\times$10$^{3}$  &  1.5$\times$10$^{5}$  & \llap{$-$}3.0$\times$10$^{5}$  &  2.9$\times$10$^{5}$ \\
$y_{16}$ & 3.5$\times$10$^{4}$  &  1.6$\times$10$^{5}$  & \llap{$-$}2.9$\times$10$^{5}$  &  3.6$\times$10$^{5}$ \\
$y_{17}$ & 6.5$\times$10$^{4}$  &  2.1$\times$10$^{5}$  & \llap{$-$}3.5$\times$10$^{5}$  &  4.8$\times$10$^{5}$ \\
$y_{18}$ & 9.9$\times$10$^{4}$  &  4.6$\times$10$^{5}$  & \llap{$-$}8.2$\times$10$^{5}$  &  1.0$\times$10$^{6}$ \\
$y_{19}$ &\llap{$-$}1.6$\times$10$^{4}$  &  1.1$\times$10$^{6}$  & \llap{$-$}2.3$\times$10$^{6}$  &  2.2$\times$10$^{6}$ \\
$y_{20}$ &\llap{$-$}2.8$\times$10$^{6}$  &  3.9$\times$10$^{6}$  & \llap{$-$}1.1$\times$10$^{7}$  &  4.9$\times$10$^{6}$ \\
$y_{21}$ & 7.3$\times$10$^{6}$  &  5.2$\times$10$^{6}$  & \llap{$-$}3.1$\times$10$^{6}$  &  1.8$\times$10$^{7}$ \\
$y_{22}$ &\llap{$-$}1.6$\times$10$^{6}$  &  5.9$\times$10$^{6}$  & \llap{$-$}1.3$\times$10$^{7}$  &  1.0$\times$10$^{7}$ \\
$y_{23}$ & 1.0$\times$10$^{7}$  &  8.9$\times$10$^{6}$  & \llap{$-$}7.6$\times$10$^{6}$  &  2.8$\times$10$^{7}$ \\
$y_{24}$ &\llap{$-$}4.0$\times$10$^{6}$  &  1.6$\times$10$^{7}$  & \llap{$-$}3.5$\times$10$^{7}$  &  2.7$\times$10$^{7}$ \\
$y_{25}$ &\llap{$-$}4.1$\times$10$^{6}$  &  2.9$\times$10$^{7}$  & \llap{$-$}6.1$\times$10$^{7}$  &  5.3$\times$10$^{7}$ \\
\bottomrule
\end{tabular}
\caption{\label{tab:coeffsB} Expectation values (second column), standard deviations (third column), 95\% CL lower (fourth column) and upper bounds (last column) for the 25 non-birefringent {\bf rotated} coefficients of the SME phton sector for $d=6$.}
\end{table}



\section{Conclusions} \label{sec:conclusions}

In this work, we tackled the lack of a clear connection between the commonly used constraints on LIV, typically expressed via the effective quantum gravity scale $E_{QG,n}$, and the coefficients of the SME in the pure photon sector. Although these bounds are widely used in the literature, they are rarely translated into the SME framework. Here, we developed a consistent and straightforward formalism to perform this conversion and showed how dispersion relations from the general LIV approach can be systematically mapped onto the SME description.

To make meaningful comparisons and enable a coherent global analysis, we carried out a critical review and standardization of some of the most stringent $E_{QG,1}$ and $E_{QG,2}$ bounds available. This included adding missing prefactors, accounting for systematic uncertainties in the energy scale, and converting one-sided bounds into two-sided Gaussian uncertainties where appropriate. The resulting methodology lays a solid foundation for expressing time-of-flight constraints in terms of SME coefficients and sets the stage for combining results from a broad range of astrophysical sources. 

The best bound, converted from LHAASO's analysis of GRB 221009A data~\citep{lhaaso_2024} leads to an improvement of bounds on $\sum_{j m}  Y_{j m} c_{(I) j m}^{(6)}$ by one and a half order of magnitude with respect to previous ones, albeit possible intrinsic energy-dependent photon time delays were not considered in their analysis. We may, hence, not exclude the possibility that the photon detection times were potentially affected by intrinsic delays in the source, and nature has conspired to compensate for a LIV-induced effect to provide a final null measurement.  It is therefore essential to expand the sample through analyses of similar or better sensitivity from a variety of astrophysical sources at different distances and of different origins to eliminate such a scenario. 

A decent statistical treatment, beyond the most straightforward approach used in~\citet{Kislat:2015}, of a large variety of measurements, many of which use a different treatment of uncertainties and (choice of) underlying systematics, has revealed challenges. Apart from the fact that none of the most sensitive analyses have incorporated a statistical treatment of intrinsic time delays or a correct marginalization of the sometimes quite considerable detector-related systematics, the standards themselves differ. We have tried, as best as we could with the available information, to standardize at least confidence intervals, detector systematics, inversion formulae and conversion of the measurement of best and asymmetric confidence intervals to expectation values and standard deviations. Our wish is to have provided a first step toward a community-wide standardization in all of these aspects.  We urge the community to publish their full likelihoods, plus expectation values and variances.

Our new bounds on the individual non-birefringent coefficients of the photon sector for $d=6$ improve over previous ones~\citep{Kislat:2015} by about an order of magnitude. This is a direct consequence of the improvement in the \textit{least sensitive} measurement in the sample of the best 25, which determines the order of magnitude of the bounds on all coefficients  $c_{(I)jm}^{(d)}$. Note that the largest standard deviation on the rotated coefficients, $s(y_{25})$, provides an upper bound to the standard deviations of the original coefficients, which themselves vary by only within a factor of $\sim 8$. 

Reversing this argument, a set of fourteen additional competitive bounds from very-high-energy or ultra-high-energy gamma-ray observatories could improve sensitivity to all $c_{(I)jm}^{(d)}$ by another five orders of magnitude!

\acknowledgments
It is a pleasure to thank Alan Kosteleck\'y and Matt Mewes for discussions and helpful suggestions. We also thank Michele Doro for helpful discussions and suggestions on the statistics part. 
M.Gu.\ and R.Po.\ acknowledge support by CIDMA under the FCT Multi-Annual Financing Program for R\&D units. M.Gu acknowledges the support in part by a fellowship by the Hanse-Wissenschaftskolleg - Institute for Advanced Study (HWK) and Horizon Europe staff exchange (SE) programme HORIZON-MSCA2021-SE-01 Grant No. NewFunFiCO-101086251. M.Ga.\ and A.Ca.\ acknowledge funding from the Spanish grant PID2022-139117NB-C43, funded by MCIN/AEI/10.13039/501100011033/FEDER, UE. 

\newpage
\appendix 

\begin{table}[h]
    \centering
    \renewcommand{\arraystretch}{1.2}
    {\small
    \begin{tabular}{lcccc}
        \toprule
        \textbf{Source Name} & \textbf{R.A.} & \textbf{Dec.}  & E[$\sum_{jm}{}_0Y_{jm}(\theta,\phi)c^{(6)}_{(I)jm}$] & $s(\sum_{jm}{}_0Y_{jm}(\theta,\phi)c^{(6)}_{(I)jm})$\\
        &  (°) &  (°) & (GeV\(^{-2}\))   & (GeV\(^{-2}\)) \\
        \midrule
            3C454.3 & 343.5  & 16.1  & 2.3$\times 10^{-19}$ &  4.3$\times 10^{-19}$ \\
 PKS1222+216 & 186.2  & 21.4  & 2.4$\times 10^{-18}$ &  7.0$\times 10^{-18}$ \\
       3C279 & 194.0  & \llap{$-$}5.8 &  2.3$\times 10^{-18}$ &  9.4$\times 10^{-18}$ \\
 PKS1502+106 & 226.1  & 10.5 &   0.6$\times 10^{-17}$ &  1.3$\times 10^{-17}$ \\
 PKS1510-089 & 228.2  & \llap{$-$}9.1 &   0.3$\times 10^{-17}$ &  1.3$\times 10^{-17}$ \\
  PKS1424-41 & 217.0  & \llap{$-$}42.1 &  \llap{$-$}0.2$\times 10^{-17}$ &  2.3$\times 10^{-17}$ \\
 PKS1830-211 & 278.4  & \llap{$-$}21.1 & \llap{$-$}0.3$\times 10^{-17}$ &  2.8$\times 10^{-17}$ \\
   S41849+67 & 282.3  & 67.1 & \llap{$-$}0.5$\times 10^{-16}$ &  1.3$\times 10^{-16}$ \\ 
       3C273 & 187.3  &  2.1 &   0.2$\times 10^{-16}$ &  1.7$\times 10^{-16}$ \\
      B21520+31 & 230.5 & 31.7 & \llap{$-$}1.8$\times 10^{-16}$   &3.3$\times 10^{-16}$ \\ 
    PKS0426-380 &  67.2 & \llap{$-$}37.9&  0.5$\times 10^{-16}$   &6.7$\times 10^{-16}$ \\ 
    PKS0716+714 & 110.5 & 71.3 &\llap{$-$}6.6$\times 10^{-16}$    &7.8$\times 10^{-16}$ \\ 
      GB1310+487&  198.2&  48.5&  7.1$\times 10^{-16}$   & 9.9$\times 10^{-16}$ \\ 
    PKS2326-502 & 352.2 &\llap{$-$}49.9 & 1.3$\times 10^{-15}$   &1.7$\times 10^{-15}$ \\ 
    PKS0454-234 &  74.3 &\llap{$-$}23.4 & 0.8$\times 10^{-15}$   &1.7$\times 10^{-15}$ \\ 
    PKS1633+382 & 248.8 & 38.1 &\llap{$-$}0.09$\times 10^{-15}$   &1.7$\times 10^{-15}$ \\ 
     B31343+451 & 206.4 & 44.9 &\llap{$-$}0.06$\times 10^{-15}$   &2.4$\times 10^{-15}$ \\ 
      S30218+35 &  35.3 & 35.9 &\llap{$-$}0.1$\times 10^{-15}$   &3.7$\times 10^{-15}$ \\ 
    PKS2233-148 & 339.1 &\llap{$-$}14.6 &\llap{$-$}0.9$\times 10^{-15}$  & 4.3$\times 10^{-15}$ \\ 
         4C14.23 & 111.3&  14.4 &\llap{$-$}0.02$\times 10^{-15}$  & 4.5$\times 10^{-15}$ \\ 
  PMNJ2345-1555  & 356.3& \llap{$-$}15.9&  0.6$\times 10^{-15}$  & 4.7$\times 10^{-15}$ \\ 
     PKS0235+164 &  39.7&  16.6 & 3.0$\times 10^{-15}$  & 6.5$\times 10^{-15}$ \\ 
         4C28.07 &  39.5&  28.8 &\llap{$-$}0.08$\times 10^{-14}$  & 1.3$\times 10^{-14}$ \\ 
           3C66A &  35.7&  43.0 & 0.5$\times 10^{-14}$  & 2.1$\times 10^{-14}$ \\
        \bottomrule
\end{tabular}}
    \caption{Approximated expectation values and standard deviations of the 24 AGNs of \citet{Kislat:2015} used to increase the data sample for the inversion of the individual coefficients $c^{(6)}_{(I)jm}$. 
    \label{tab:LIV_table_AGNs} }
\end{table}

\begin{table}[h]
    \centering
    \renewcommand{\arraystretch}{1.2}
    {\scriptsize
    \begin{tabular}{lccccc}
        \toprule
        \textbf{Source Name} & \textbf{R.A.} & \textbf{Dec.} & \(\sum_{jm}{}_0Y_{jm}(\theta,\phi)c^{(6)}_{(I)jm}\) & E[$\sum_{jm}{}_0Y_{jm}(\theta,\phi)c^{(6)}_{(I)jm}$] & $s(\sum_{jm}{}_0Y_{jm}(\theta,\phi)c^{(6)}_{(I)jm})$\\
        &  (°) &  (°) & (GeV\(^{-2}\))  & (GeV\(^{-2}\)) & (GeV\(^{-2}\)) \\
        \midrule
        GRB 210619B  & 319.7  & $+33.9$  & \(1.20^{+0.45}_{-0.46} \times 10^{-15}\) & \llap{1}$2.0\times 10^{-16}$ & $2.3\times 10^{-16}$\\
        GRB 160625B$\,^*$  & 308.6  & $+6.9$   & 
        & \(1.6 \times 10^{-15}\) &  
        \(1.4 \times 10^{-15}\)\rlap{\,$^\dagger$}
        \\
        GRB 180720B  & 0.59   & $-3.0$   & \llap{$-$}\(0.03^{+0.70}_{-0.65} \times 10^{-14}\) &
        \llap{$-$}\(0.1 \times 10^{-15}\)
        & \(3.4 \times 10^{-15}\) \\
        GRB 200829A  & 251.1  & $+72.4$  & \(3.3^{+0.9}_{-1.1} \times 10^{-14}\) & 
        \llap{3}\(1.8 \times 10^{-15}\)
        & \(4.9 \times 10^{-15}\)
        \\
        GRB 130427A  & 173.1  & $+27.7$  & \(4.7^{+5.7}_{-6.1} \times 10^{-14}\) & 
        \(4.5 \times 10^{-14}\) & 
        \(2.9 \times 10^{-14}\)\\
        GRB 130518A  & 355.7  & $+47.5$  & \(7.0^{+6.3}_{-5.8} \times 10^{-14}\) & \(7.2 \times 10^{-14}\) & \(3.0\times 10^{-14}\) \\
        GRB 150314A  & 126.7  & $+63.8$  & \(1.3^{+2.4}_{-2.7} \times 10^{-13}\) & 
        \(1.1 \times 10^{-13}\)
        & 
        \(1.3 \times 10^{-13}\)
        \\
        GRB 091003A  & 251.5  & $-36.6$  & \(2.7^{+3.7}_{-2.3} \times 10^{-13}\) & \(3.4 \times 10^{-13}\) & \(1.5 \times 10^{-13}\) \\
        GRB 131108A  & 156.5  & $+9.7$  & \(0.1^{+4.5}_{-2.9} \times 10^{-13}\)  & \(0.9 \times 10^{-13}\)   & \(1.8 \times 10^{-13}\) \\
        GRB 150403A  & 311.5  & $-62.7$  & \(7.1^{+8.9}_{-8.0} \times 10^{-13}\) & \(7.6 \times 10^{-13}\) & \(4.2 \times 10^{-13}\) \\
        GRB 140206A  & 145.3  & $+66.8$  & \(0.9^{+1.0}_{-0.8} \times 10^{-12}\) & \(9.8 \times 10^{-13}\) & \(4.5 \times 10^{-13}\) \\
        GRB 160509A  & 310.1  & $+76.0$  & \(5.6^{+11.9}_{-10.9} \times 10^{-13}\) & \(6.1 \times 10^{-13}\) & \(5.7 \times 10^{-13}\) \\
        GRB 140508A  & 255.5  & $+46.8$  & \(1.7^{+1.6}_{-1.5} \times 10^{-12}\) & \llap{1}\(7.4 \times 10^{-13}\) & \(7.8 \times 10^{-13}\) \\
        GRB 201216C  & 16.4   & $+16.5$  & \(0.9^{+2.5}_{-2.5} \times 10^{-12}\) & \(0.9 \times 10^{-12}\) & \(1.3 \times 10^{-12}\) \\
        GRB 141028A  & 322.6  & $-0.2$   & \(2.1^{+4.4}_{-2.8} \times 10^{-12}\)  & \(3.0 \times 10^{-12}\)   & \(1.8\times 10^{-12}\) \\
        GRB 120119A  & 120.0  & $-9.8$   & \(3.1^{+3.9}_{-3.3} \times 10^{-12}\) & \(3.4 \times 10^{-12}\) & \(1.8 \times 10^{-12}\) \\
        GRB 131231A  & 10.6   & $-1.6$   & \(1.04^{+0.42}_{-0.44} \times 10^{-11}\)  & \llap{1}\(0.3 \times 10^{-12}\)  & \(2.2 \times 10^{-12}\)  \\
        GRB 180703A  & 6.5    & $-67.1$  & \(2.4^{+7.4}_{-3.3} \times 10^{-12}\) & \(4.5 \times 10^{-12}\) & \(2.7 \times 10^{-12}\) \\
        GRB 081221   & 15.8   & $-24.5$  & \(6.5^{+6.9}_{-5.6} \times 10^{-12}\) & \(7.2 \times 10^{-12}\) & \(3.1 \times 10^{-12}\) \\
        GRB 090328   & 155.7  & $+33.4$  & \(7.5^{+6.8}_{-6.7} \times 10^{-12}\) & \(7.5 \times 10^{-12}\)  & \(3.4 \times 10^{-12}\) \\
        GRB 200613A  & 153.0  & $+45.8$  & \(0.59^{+0.84}_{-0.72} \times 10^{-11}\) & \(6.5 \times 10^{-12}\) & \(3.9\times 10^{-12}\) \\
        GRB 100728A  & 88.8   & $-15.3$  & \llap{$-$}\(0.7^{+10.3}_{-5.8} \times 10^{-12}\) & \(1.5 \times 10^{-12}\) & \(4.0 \times 10^{-12}\) \\
        GRB 210610B  & 243.9  & $+14.4$  & \llap{$-$}\(0.8^{+10.8}_{-8.4} \times 10^{-12}\) & \(0.4\times 10^{-12}\) & \(4.8 \times 10^{-12}\) \\
        GRB 210204A  & 109.1  & $+9.7$   & \(0.2^{+11.9}_{-8.9} \times 10^{-12}\) & \(1.7 \times 10^{-12}\)  & \(5.2\times 10^{-12}\) \\
        GRB 090618   & 294.0  & $+78.4$  & \(1.5^{+1.4}_{-1.3} \times 10^{-11}\)  & \llap{1}\(5.8 \times 10^{-12}\)  & \(6.7 \times 10^{-12}\) \\
        GRB 150514A  & 74.8   & $-60.9$  & \(1.6^{+2.4}_{-1.6} \times 10^{-11}\)  & \(2.0\times 10^{-11}\)   & \(1.0 \times 10^{-11}\)  \\
        GRB 171010A  & 66.6   & $-10.5$  & \(3.5^{+4.1}_{-12.5} \times 10^{-11}\) & \llap{$-$}\(0.7 \times 10^{-11}\) & \(4.1 \times 10^{-11}\) \\
        GRB 150821A  & 341.9  & $-57.9$  & \(0.7^{+1.2}_{-0.6} \times 10^{-10}\)  & \(9.5 \times 10^{-11}\)  & \(4.4 \times 10^{-11}\) \\
        GRB 130925A  & 41.2   & $-26.1$  & \(2.9^{+12.4}_{-7.0} \times 10^{-11}\) & \(5.6 \times 10^{-11}\) & \(4.8 \times 10^{-11}\) \\
        \bottomrule
    \end{tabular}}
    \caption{Extracted values of \(\sum_{jm}{}_0Y_{jm}(\theta,\phi)c^{(6)}_{(I)jm}\) for those 29 GRBs of \citet{Wei:2022} 
    used to increase the data sample for the inversion of the individual coefficients $c^{(6)}_{(I)jm}$. In the last-but-one and last columns, an approximation of the expectation value and standard deviation has been made, following the presciption of \citet{DAgostini:2004}.
    $^*$~Due to the apparently problematic fit of the spectral lags of this GRB to the data in Fig.~1 of~\citet{Wei:2022} (without the corresponding fit $\chi^2/\mathrm{ndf}$ provided), we preferred to derive expectation value and standard deviation of GRB~160625B from the upper bounds provided in~\citet{Wei:2017}.
    $^\dagger$~The standard deviation has been corrected with the square root of the fit  $\chi^2/\mathrm{ndf}$.
    \label{tab:LIV_table_GRBs} }
\end{table}

\begin{equation}
\rotatebox{90}{$%
Q_{ij} = \left(
{\tiny 
\begin{smallmatrix}
  0.5 &  0.06 &  0.1 &  0.4 &  -0.4 &  -0.1 &  -0.08 &  0.02 &  0.3 &  0.2 &  -0.2 &  -0.1 &  0.05 &  -0.1 &  -0.3 &  0.2 &  -0.07 &  0.03 &  0.08 &  0.1 &  -0.1 &  0.05 &  0.004 &  -0.06 &  -0.1\\
  -0.2 &  -0.4 &  0.3 &  -0.01 &  -0.2 &  0.2 &  -0.2 &  -0.3 &  0.2 &  0.1 &  0.2 &  -0.3 &  0.1 &  0.08 &  -0.07 &  -0.3 &  0.3 &  0.06 &  0.1 &  -0.06 &  0.1 &  0.01 &  0.007 &  0.09 &  0.04\\
  -0.2 &  0.1 &  0.2 &  -0.05 &  0.2 &  0.09 &  -0.1 &  -0.2 &  0.4 &  0.1 &  -0.4 &  0.3 &  0.04 &  0.3 &  -0.1 &  -0.3 &  -0.3 &  -0.09 &  -0.03 &  0.04 &  -0.2 &  -0.1 &  -0.03 &  0.1 &  0.01\\
  -0.07 &  0.1 &  -0.1 &  -0.01 &  0.04 &  0.005 &  -0.3 &  -0.03 &  -0.3 &  0.5 &  0.2 &  0.009 &  0.08 &  -0.2 &  -0.04 &  -0.2 &  -0.2 &  0.2 &  0.07 &  0.4 &  0.06 &  0.1 &  -0.02 &  0.3 &  -0.1\\
  -0.3 &  -0.2 &  -0.06 &  -0.08 &  -0.4 &  -0.03 &  -0.3 &  0.2 &  -0.01 &  -0.3 &  -0.07 &  -0.1 &  -0.2 &  0.2 &  -0.1 &  0.1 &  -0.3 &  -0.1 &  -0.3 &  0.3 &  0.1 &  -0.1 &  -0.1 &  0.04 &  -0.1\\
  -0.03 &  0.2 &  -0.2 &  0.1 &  -0.1 &  -0.05 &  -0.1 &  -0.2 &  -0.09 &  -0.3 &  -0.08 &  0.06 &  0.3 &  -0.08 &  0.03 &  -0.2 &  0.4 &  -0.07 &  -0.3 &  0.2 &  -0.3 &  -0.2 &  0.4 &  0.02 &  -0.1\\
  -0.2 &  -0.3 &  0.08 &  0.1 &  -0.4 &  0.1 &  0.1 &  -0.06 &  -0.08 &  -0.1 &  0.06 &  0.4 &  0.08 &  -0.3 &  0.09 &  0.008 &  -0.3 &  0.3 &  0.1 &  -0.2 &  -0.3 &  0.08 &  0.09 &  -0.09 &  0.07\\
  0.1 &  -0.01 &  -0.2 &  0.06 &  0.02 &  0.4 &  0.2 &  -0.05 &  -0.1 &  -0.1 &  0.2 &  0.1 &  0.01 &  0.1 &  -0.4 &  -0.06 &  -0.1 &  -0.1 &  0.1 &  -0.2 &  0.1 &  -0.1 &  -0.03 &  0.1 &  -0.5\\
  -0.2 &  0.3 &  -0.3 &  0.09 &  -0.2 &  0.1 &  -0.2 &  -0.1 &  -0.09 &  0.02 &  -0.3 &  -0.2 &  0.2 &  0.1 &  -0.1 &  -0.002 &  -0.05 &  0.09 &  -0.05 &  -0.4 &  0.3 &  0.4 &  0.08 &  -0.1 &  0.1\\
  -0.005 &  -0.2 &  0.02 &  0.2 &  0.2 &  -0.4 &  -0.002 &  0.04 &  -0.1 &  -0.4 &  0.007 &  -0.3 &  0.02 &  -0.01 &  -0.09 &  -0.4 &  -0.3 &  -0.2 &  0.4 &  0.08 &  -0.01 &  0.2 &  0.2 &  -0.2 &  -0.02\\
  0.3 &  -0.4 &  -0.1 &  -0.4 &  0.02 &  -0.1 &  0.2 &  -0.2 &  -0.2 &  0.1 &  -0.1 &  -0.1 &  0.4 &  0.1 &  -0.08 &  0.1 &  -0.2 &  -0.08 &  -0.2 &  -0.02 &  -0.2 &  0.2 &  -0.01 &  0.1 &  0.07\\
  0.2 &  -0.03 &  0.2 &  0.03 &  -0.09 &  -0.2 &  0.1 &  0.08 &  -0.2 &  0.2 &  0.07 &  0.09 &  -0.2 &  0.5 &  -0.03 &  -0.2 &  -0.004 &  0.3 &  -0.2 &  -0.1 &  0.1 &  -0.07 &  0.4 &  -0.1 &  -0.04\\
  -0.2 &  0.03 &  -0.2 &  0.1 &  0.2 &  -0.08 &  0.1 &  -0.4 &  0.2 &  0.05 &  0.3 &  -0.04 &  -0.08 &  0.1 &  -0.3 &  0.3 &  -0.1 &  0.2 &  -0.1 &  0.3 &  -0.06 &  0.02 &  0.06 &  -0.4 &  0.2\\
  -0.2 &  -0.1 &  0.3 &  0.2 &  0.1 &  0.2 &  0.04 &  -0.1 &  -0.3 &  0.2 &  -0.2 &  -0.04 &  -0.3 &  -0.08 &  0.05 &  0.2 &  0.1 &  -0.3 &  -0.1 &  0.1 &  -0.2 &  0.3 &  0.1 &  -0.2 &  -0.2\\
  0.3 &  -0.4 &  -0.1 &  0.2 &  0.4 &  0.1 &  -0.4 &  0.05 &  0.01 &  -0.2 &  -0.1 &  0.3 &  -0.01 &  -0.09 &  -0.08 &  0.06 &  0.08 &  0.2 &  -0.2 &  -0.07 &  0.2 &  0.1 &  -0.004 &  0.02 &  0.04\\
  -0.05 &  -0.05 &  0.1 &  0.2 &  -0.07 &  -0.4 &  -0.1 &  -0.2 &  0.08 &  0.1 &  0.3 &  0.3 &  0.1 &  -0.2 &  0.07 &  0.1 &  -0.2 &  -0.5 &  -0.2 &  -0.3 &  0.4 &  -0.08 &  0.1 &  0.1 &  -0.07\\
  -0.2 &  -0.2 &  -0.1 &  0.2 &  0.03 &  -0.05 &  0.2 &  -0.005 &  -0.06 &  0.03 &  -0.3 &  -0.01 &  -0.1 &  0.09 &  -0.1 &  0.3 &  0.07 &  0.05 &  0.3 &  0.09 &  0.1 &  -0.2 &  0.3 &  0.6 &  0.2\\
  -0.2 &  -0.07 &  -0.2 &  0.2 &  0.2 &  -0.02 &  -0.0008 &  0.2 &  0.3 &  0.2 &  0.1 &  -0.4 &  0.05 &  0.06 &  0.3 &  0.06 &  -0.2 &  0.1 &  -0.2 &  -0.3 &  -0.3 &  0.004 &  0.1 &  0.2 &  -0.3\\
  0.2 &  0.2 &  0.4 &  0.1 &  0.06 &  0.4 &  0.06 &  0.06 &  0.008 &  -0.2 &  0.2 &  -0.09 &  0.3 &  0.1 &  0.2 &  0.2 &  -0.3 &  0.01 &  -0.09 &  0.2 &  0.2 &  0.1 &  0.2 &  0.2 &  0.2\\
  0.0005 &  -0.07 &  0.05 &  -0.06 &  0.1 &  0.01 &  0.2 &  -0.3 &  0.03 &  -0.03 &  -0.3 &  -0.2 &  0.04 &  -0.3 &  0.3 &  0.04 &  -0.2 &  0.3 &  -0.02 &  0.1 &  0.4 &  -0.4 &  0.01 &  -0.2 &  -0.3\\
  0.3 &  -0.1 &  -0.3 &  0.3 &  -0.1 &  0.2 &  -0.1 &  -0.3 &  -0.2 &  0.1 &  -0.008 &  0.02 &  -0.1 &  0.3 &  0.4 &  -0.1 &  -0.1 &  -0.2 &  0.1 &  0.03 &  -0.07 &  -0.2 &  -0.2 &  -0.1 &  0.2\\
  -0.1 &  -0.2 &  -0.03 &  -0.2 &  0.07 &  0.1 &  -0.3 &  0.3 &  -0.02 &  0.2 &  -0.02 &  0.09 &  0.4 &  0.2 &  -0.02 &  0.3 &  0.004 &  -0.1 &  0.3 &  0.01 &  -0.002 &  -0.2 &  0.3 &  -0.4 &  -0.04\\
  0.08 &  -0.2 &  -0.3 &  -0.06 &  -0.09 &  0.2 &  0.3 &  0.2 &  0.4 &  0.2 &  -0.06 &  0.2 &  -0.06 &  -0.1 &  0.1 &  -0.3 &  0.01 &  -0.2 &  -0.08 &  0.3 &  0.2 &  0.3 &  0.3 &  -0.1 &  0.07\\
  -0.2 &  -0.08 &  0.05 &  0.3 &  -0.02 &  -0.2 &  0.2 &  0.07 &  0.009 &  0.01 &  -0.05 &  0.2 &  0.4 &  0.3 &  0.2 &  0.04 &  0.2 &  0.1 &  0.06 &  0.2 &  0.1 &  0.3 &  -0.4 &  -0.0002 &  -0.3\\
  -0.1 &  -0.09 &  0.1 &  0.3 &  0.1 &  0.08 &  0.2 &  0.3 &  -0.2 &  0.2 &  -0.04 &  -0.1 &  0.3 &  -0.2 &  -0.3 &  -0.2 &  -0.001 &  -0.1 &  -0.3 &  -0.05 &  0.04 &  -0.3 &  -0.2 &  -0.1 &  0.3\\
\end{smallmatrix}} 
\right)$}
\label{eq:Qij}
\end{equation}

\newpage
\bibliographystyle{elsarticle-harv}
\bibliography{bibliography}

\end{document}